\documentclass[11pt,twoside]{article} 
\usepackage{asp2004}
\usepackage{graphicx}
\usepackage{verbatim}
\usepackage{lscape} 
\usepackage{natbib}

\begin{document} 
\title{Alkali Line Profiles in Ultracool White Dwarfs}
\author{Derek Homeier,$^1$ 
        Nicole Allard,$^2$  
        Christine M. S. Johnas,$^3$ Peter H. Hauschildt,$^3$ and 
        France Allard$^{4,5,6}$}  
\affil{$^1$ Institut f{\"u}r Astrophysik, Georg-August-Universit{\"a}t, 
  Friedrich-Hund-Platz 1, 37077 G{\"o}ttingen, Germany\\ 
$^2$Institut d'Astrophysique de Paris, CNRS,  
  98bis Boulevard Arago, 75014 Paris, France\\ 
$^3$ Hamburger Sternwarte, Gojenbergsweg\,112, 21029\,Hamburg,~Germany\\ 
$^4$ Universit\'e de Lyon, 69003 Lyon, France\\ 
$^5$ Ecole Normale Sup\'erieure de 
  Lyon, 46 all\'ee d'Italie, 69007 Lyon, France\\ 
  $^6$ CNRS, UMR 5574, Centre de Recherche Astrophysique de Lyon, 
  Universit\'e Lyon~1, 69622 Villeurbanne, France} 

\begin{abstract} 
We present \texttt{PHOENIX} atmosphere models for metal-rich 
cool white dwarfs using improved line shapes for the Na\,I and K\,I
resonance doublets.  
Profiles for collisional broadening due 
to H$_2$ and He based on the adiabatic representation show strong
deviations from Van der Waals interaction at short distances. 
Comparison with observed spectra that 
show extremely broadened Na\,I lines 
indicates that a He-rich atmospheric composition is required
to explain the line strengths and spectral energy distributions. 
Our current synthetic spectra, using an 
expansion in powers of density to the third order optimised for brown
dwarf atmosphere conditions, significantly underestimate the observed
absorption in the far wings, even predicting smaller total line
strength than a Lorentzian profile. 
This is shown to be due to the handling of multiple perturber
interactions becoming inadequate for the extreme densities of
the coolest white dwarfs. 
The density expansion would have to be extended at least to
the 7$^\mathrm{th}$ order for an accurate treatment of such conditions 
and might break down altogether in the densest objects. 
The results of a direct calculation of the unified
profile should therefore be used for model atmospheres of cool
metal-rich white dwarfs. 
Qualitative comparison of the full adiabatic profile to the spectrum of
WD2356$-$209  indicates good agreement with the observed line shape.  
Observations of the coolest white
dwarfs may therefore serve as a laboratory for testing the physics of
the deeper atmospheres and interiors of brown dwarfs and giant
planets. 

\end{abstract}

\section{Introduction}
Observations of the oldest and coldest white dwarfs following
the study of \citet{OHDHS01} have revealed 
two stars showing very unusual wide and deep absorption at
5000\,--\,6000\,{\AA}.  
In WD2356$-$209 this spectroscopic feature is accompanied by 
a unusual combination of blue {\bv} and extremely red
{\hbox{$V\!-\!I$}} colours \citep{SalimHalo04}.
A second star with this absorption feature has been identified in Data
Release 2 of the Sloan Digital Sky Survey: 
SDSS\,J\,13\,30\,01.13+64\,35\,23.8 \citep[hereafter
SDSS\,J1330+6435][]{sdssWD03}, which also shows indications of an
additional strong absorption around 4000\,{\AA}.  
These spectra bear intriguing resemblance to the strongly broadened 
Na\,I and K\,I resonance doublets observed in cool brown dwarfs of late L
or T spectral type \citep{tsujiGl229,BurrowsMS00}. 
Collisional broadening of the Na\,I lines in a metal-enriched,
extremely dense white dwarf atmosphere thus lent itself as an 
explanation for these observations. 

\begin{figure}[htbp]
  \centering
  \includegraphics[width=0.9\hsize,clip=true,trim=3 3 10 20]{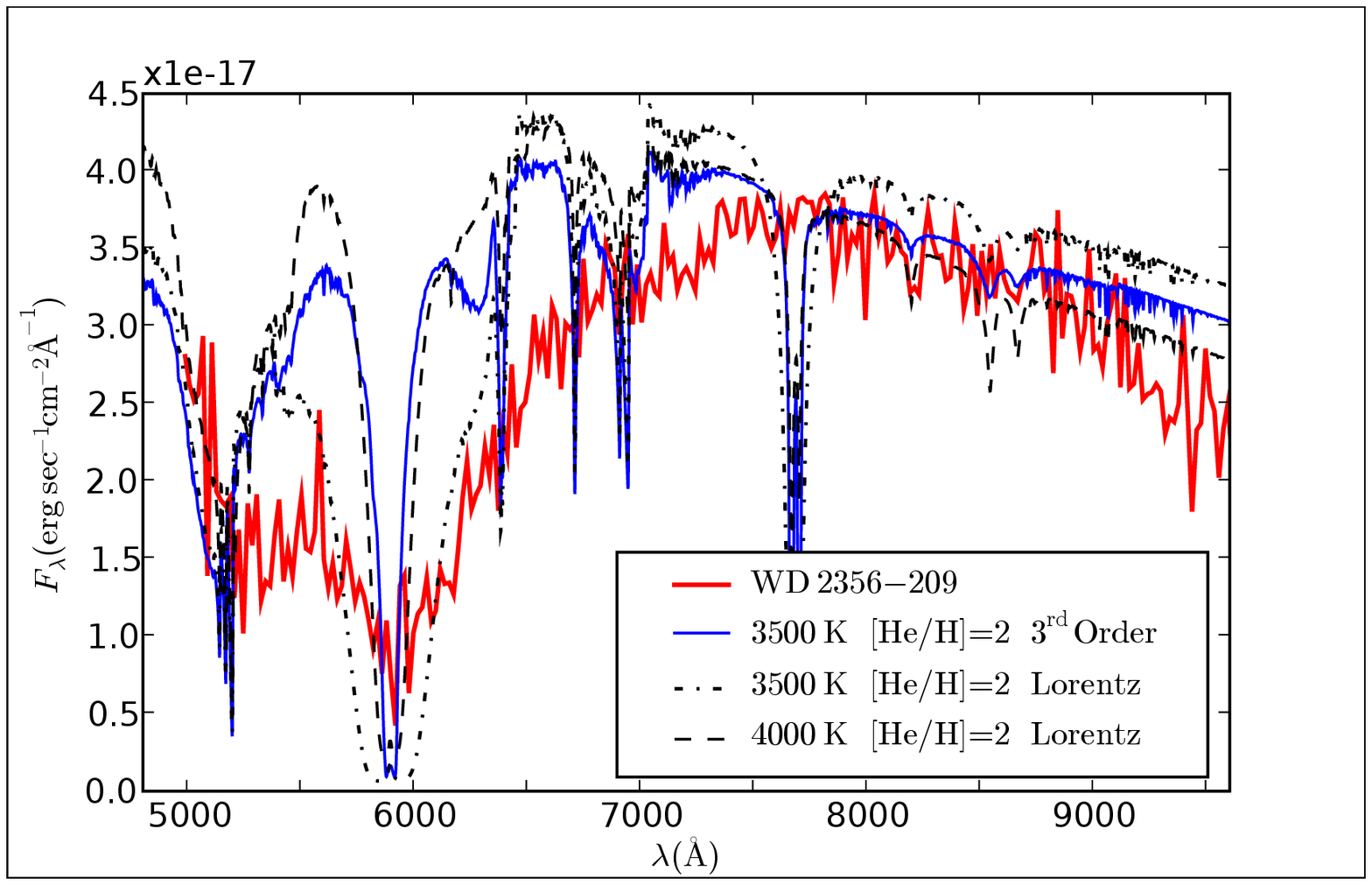}
  \vspace{-0.8ex}
  \caption{Comparison of the observed spectrum of WD2356$-$209
    \citep{OHDHS01} to \texttt{PHOENIX} spectra using different models
    of the Na\,I line profile. The first one (smooth solid line)
    is calculated from the unified theory of \citet{NicolePhysRev99}
    in its expansion to the third power in density, the other two are
    using a Lorentzian line shape with the half-widths given
    by \citet{Cores06}. }
  \vspace{-1.2ex}
  \label{fig:wd2356}
\end{figure}

In addition to the Na\,I line the spectrum of WD2356$-$209 shows
some absorption features indicating the possible presence of other
opacity sources. In a previous analysis we have attempted to model 
these features with molecular bands expected to form
in a cool hydrogen-rich environment \citep{alkaliWDkiel}. 
In particular, chemical equilibrium models predict MgH and CaH
absorption around 5000 and 7000\,{\AA}, respectively, for effective
temperatures below approximately 5000\,K. 
However $S/N$ and resolution of the spectra would not allow the
identification of these bands with certainty. 
On the other hand strong arguments exist in favour of a helium-rich
composition, the most important ones being the generally higher
densities in the highly transparent cool helium atmospheres, which
would easily provide the extreme perturber densities evident in the
line profile, and much shorter diffusion timescales in hydrogen
atmospheres \citep{dupuis93b}, making the presence of significant 
amounts of heavier elements practically impossible without a
continuous source of enrichment. 

Yet under these premises a helium-dominated atmosphere containing
a smaller fraction of hydrogen-rich material could still provide a viable
explanation of the observed spectra. 
\citet{detlevDAZB05} have analysed an, albeit hotter, WD of such
composition 
and found its combination of hydrogen lines with an unusually strong
Balmer decrement and Ca, Mg and Fe lines best explained with the 
admixture of ca.\ 10$^{-2.8}$ solar composition material to a helium
atmosphere. Such a model also obviates the need for removing the
hydrogen from the accreted material e.\,g.\ by means of the propeller
mechanism \citep{WT82}. 

\begin{figure}[htbp]
  \centering
  \includegraphics[width=0.9\hsize,clip=true,trim=8 8 8 18]{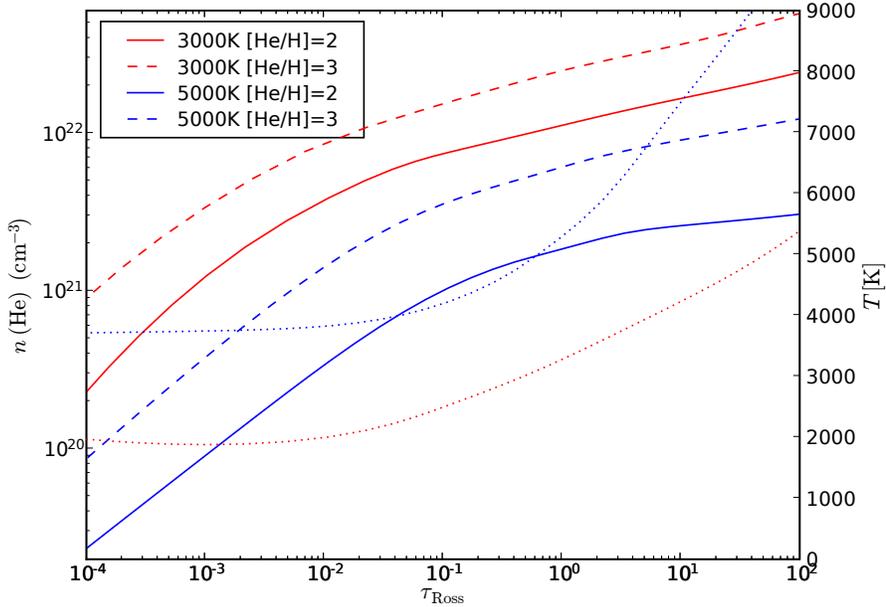}
  \vspace{-2.4ex}
  \caption{Number density of helium (left scale) and temperature
    profile (right scale) for 3000\,K (lower curves) and 5000\,K 
    WD atmosphere models with varying H/He ratio}
  \label{fig:HeHdens}
  \vspace{-1.8ex}
\end{figure}

\section{Spectral Models}
\begin{figure}[htbp]
  \centering
  \includegraphics[width=0.8\hsize,clip=true,trim=0 0 0 46]{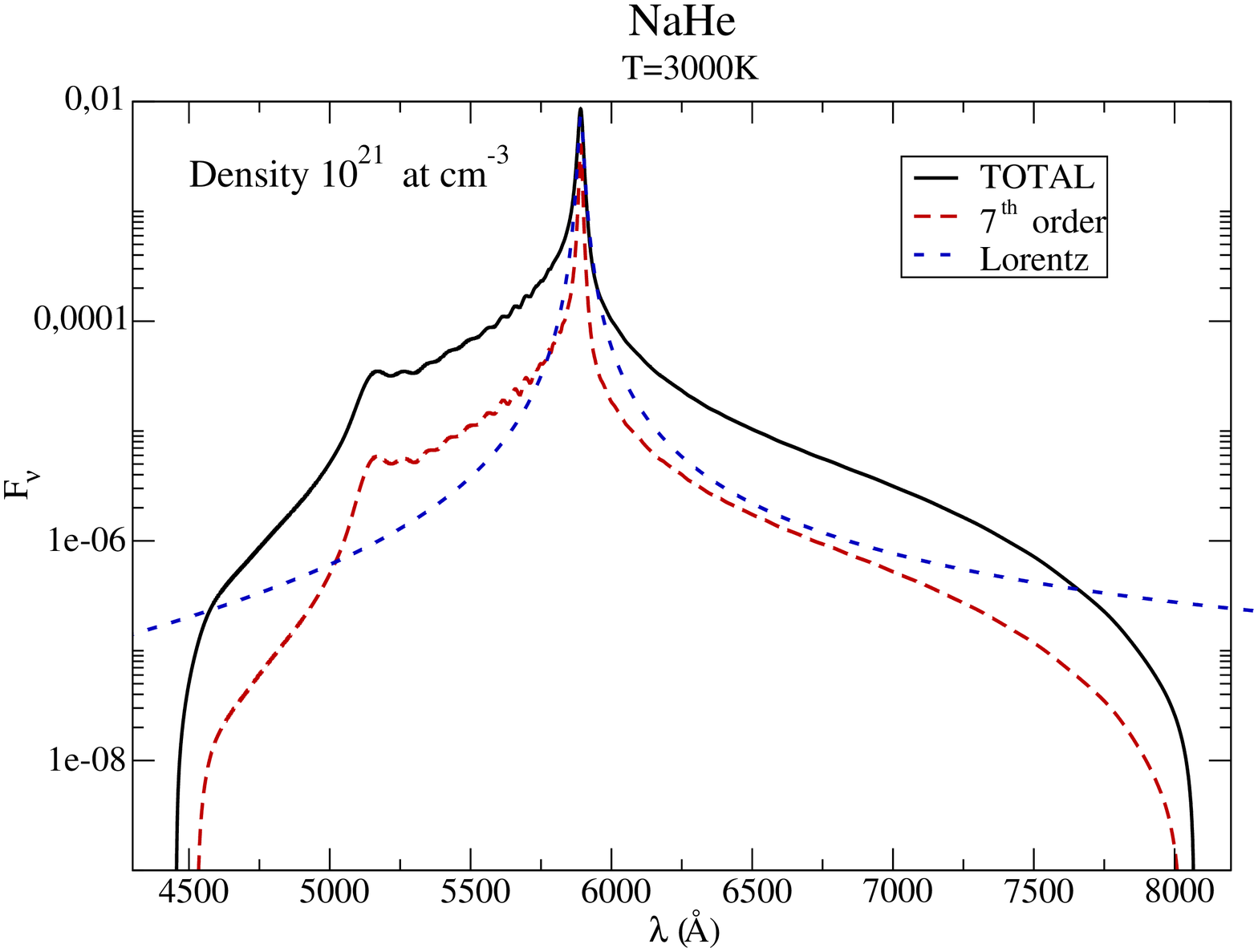}
  \vspace{-1.2ex}
  \caption{Full unified profile of Na\,I\,D2 perturbed
    by 10$^{21}\,\mathrm{cm}^{-3}$ of He at 3000\,K, compared to
    the 7$^\mathrm{th}$ order expansion and Lorentzian}
  \vspace{-2.0ex}
  \label{fig:NaHe21p7}
\end{figure}

Stellar atmosphere models were calculated using the 
\texttt{PHOENIX} code \citep{hbjcam99}, in its 
static, 1D LTE setup. 
Formation of molecules in full chemical equilibrium is 
included by the equation of state of \citet{LimDust} 
as described in \citet{alkaliWDkiel}, though  
in the white dwarf models chemistry is dominated by only a few
hydrogen compounds. 
This EOS does not take into account, on the other hand, non-ideal-gas
effects relevant at extreme densities. But densities in the line
formation region mostly do not exceed the limit of
$\sim$10$^{22}$\,cm$^{-3}$ (cf.\ Fig.\ \ref{fig:NaHe21p7}), 
where a fluid-like state is approached and corrections for
non-ideality must be included 
\citep{2005ASPC..334..203K}. 

Profiles of the alkali metal resonance lines for perturbations due to
neutral He and H$_2$ are calculated as the Fourier transform of the
autocorrelation function of the dipole moment within the adiabatic 
theory, as detailed in \citet{Alkalis03,alkalisLi}. 
In the \texttt{PHOENIX} implementation the line opacity is calculated  
by splitting the profile into a core component, which is describing 
the interactions at long distance with a Lorentzian line
shape, and the far wings for close interactions that can produce 
detunings up to several 1000\,{\AA}. 
The latter is computed in the low density limit using an expansion of
the autocorrelation function in powers of density, which for
the present models has been developed to the third order and
evaluated at a perturber density of 10$^{20}$\,cm$^{-3}$. 
This method, successfully applied under the conditions of brown dwarf
atmospheres, possibly compromises results at higher density
ranges, where multiple perturber interactions become important even at
close distances.  

\subsection{Comparison with Observation}
Figure \ref{fig:wd2356} shows the optical spectrum of WD2356$-$209 in
comparison to three of our models. In all cases shown here, the 
logarithmic ratio of He/H number densities is set to $-$2, 
i.\,e.\ $\sim$\,1\,\% hydrogen with a corresponding amount of heavy 
elements at solar-composition ratios added to the helium atmosphere . 

A comparison of the WD2356$-$209 spectrum with hydrogen/helium LTE
atmospheres is presented in Fig.~\ref{fig:wd2356}. The strong
dependence of the alkali lines on temperature, becoming quickly weaker
with increasing $T_\mathrm{eff}$, is evident. 
For these models the spectrum becomes increasingly difficult to reproduce
at temperatures above 4000\,K, as the the width of the Na\,I doublet
would require unrealistically high Na abundances, even up to the solar
value. 
Such high metallicities would also conflict with the observed limits 
on molecular bands such as MgH and CaH 
and the K\,I doublet at 7700\,{\AA}, all of which should be seen quite
clearly if a solar abundance pattern is assumed. 

\subsection{Influence of the Line Profiles}
The synthetic spectra using line profiles calculated in 
the impact approximation, thus including multiple-perturber interactions
and yielding a Lorentzian lineshape, show that this simpler model 
can better reproduce the total strength and width of the Na\,I line,
though the shape appears very poorly matched. 
One might therefore suspect that the expansion of the autocorrelation
function has to be extended to higher powers of density for a correct
implementation of the unified profile. 
As illustrated in Figure \ref{fig:HeHdens} the perturber density in the
line-forming region of an ultracool WD atmosphere can indeed exceed 
considerably the value of 10$^{20}\,\mathrm{cm}^{-3}$ for which the
present version of the far wing profiles was evaluated. 

This is confirmed by directly comparing the line strengths for higher
densities. However Figure \ref{fig:NaHe21p7} shows that at a He
density of 10$^{21}\,\mathrm{cm}^{-3}$ even the 7$^\mathrm{th}$ order
of the expansion falls short of the correct line strength in the wings
at $\sim$\,100\,--\,2000\,{\AA} by almost an order of magnitude. This
plot also illustrates how the unified profile approaches the impact 
approximation close to the line core as discussed by \citet{Cores06}. 
At larger separations, i.\,e.\ for closer interactions, the deviations
become severe, though the latter model at most wavelengths still
matches more closely the correct line strength than the density
expansion, especially in the red wing. At a few 100\,{\AA} separation
we find the third order expansion to underestimate the absorption by a
factor of more than 10, the 7$^\mathrm{th}$ order still by about 4,
but the Lorentzian profile by only 1\,$-$\,3. 
We can thus appreciate why the latter produces relatively better
agreement with the observation. 

\begin{figure}[htbp]
  \centering
  \includegraphics[width=0.8\hsize,clip=true,trim=0 0 0 46]{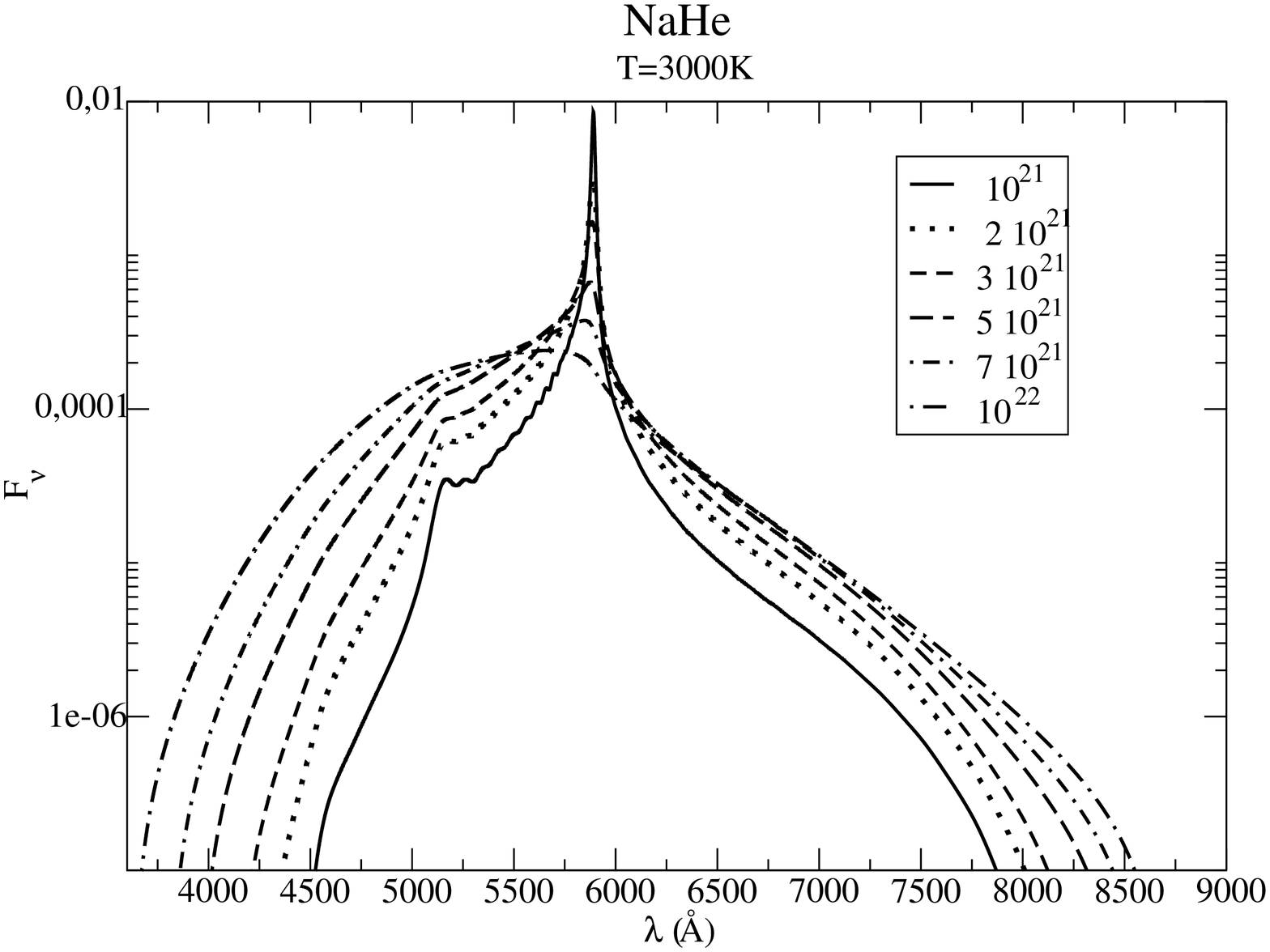}
  \vspace{-1.2ex}
  \caption{Full unified profiles of the Na\,I~D2 line for He perturber
    densities from 
    10$^{21}\,\mathrm{cm}^{-3}$ to  10$^{22}\,\mathrm{cm}^{-3}$, all
    calculations at $T$\,$=$\,3000\,K 
    }
  \vspace{-2.0ex}
  \label{fig:NaHeD2}
\end{figure}

In Figure \ref{fig:NaHeD2} we follow the further transformation of the
line profile when proceeding to yet higher densities.  
Ultimately a merging of the line core and the satellite to the blue of
the original line position emerges, increasing the opacity around
5000\,{\AA} by another factor of 10, while the red wing tends to 
asymptotically constant absorption which is actually fairly closely
matched by the Lorentz profile (not shown). 
With these simulations we may now identify much, if not all, absorption 
shortward of the Na\,I line centre with the blue wing of the resonance
line itself. Noting the sharp drop-off of opacity below 5000\,{\AA} we
can also expect that not much of the profile is formed at He densities
above 2\,$-$\,$3\times 10^{21}\,\mathrm{cm}^{-3}$. It is tempting to
trace the local flux minimum near 5000\,{\AA} to the contribution of
the satellite feature at higher layers, which is still well defined up
to 10$^{21}\,\mathrm{cm}^{-3}$, but an unambiguous identification should
require both a full model atmosphere calculation incorporating these
high density line shapes and observations at higher $S/N$. 

\section{Discussion}
We conclude that WD2356$-$209 shows the effects of collisional
damping at significantly higher perturber densities than
encountered in the atmospheres of brown dwarfs. Detailed modelling of
this object pushes broadening theory to new limits and thus provides a 
more stringent test of our calculations. This should advance our
understanding of the physical conditions in ultracool, compact objects
in general, and improve atmosphere models for brown dwarfs and other
low-mass objects as well. Although these do not show such extreme
broadening in the observable spectrum, the alkali lines provide
significant opacity in the deeper layers of their atmospheres, thereby 
affecting the thermal structure at the convective boundary and 
possibly their cooling rate (I.~Baraffe, {\em priv.\ comm.}). 
Study of ultracool white dwarfs like WD2356$-$209 or SDSS\,J1330+6435
could thereby prove a test case for modelling the evolution
of low-mass stars, brown dwarfs and giant planets. 

\acknowledgements{
  D.~H.\ acknowledges support for attending the 15th European White
  Dwarf workshop,  where this contribution was presented, under a 
  travel grant from the Deutsche Forschungsgemeinschaft (DFG) under
  number \mbox{KON 1082/2006, HO 2305/3-1}. Atmosphere  
  models presented in this work are based in part on calculations
  performed at the Gesellschaft f{\"u}r Wissenschaftliche
  Datenverarbeitung G{\"o}ttingen.}


\begin{thebibliography}{}
\expandafter\ifx\csname natexlab\endcsname\relax\def\natexlab#1{#1}\fi

\bibitem[{{Allard} {et~al.}(2001){Allard}, {Hauschildt}, {Alexander},
  {Tamanai}, \& {Schweitzer}}]{LimDust}
{Allard}, F., {Hauschildt}, P., {Alexander}, D., {Tamanai}, A., \&
  {Schweitzer}, A. 2001, \apj, 556, 357

\bibitem[{{Allard} {et~al.}(2003){Allard}, {Allard}, {Hauschildt}, {Kielkopf},
  \& {Machin}}]{Alkalis03}
{Allard}, N.~F., {Allard}, F., {Hauschildt}, P.~H., {Kielkopf}, J.~F., \&
  {Machin}, L. 2003, \aap, 411, L473

\bibitem[{{Allard} {et~al.}(2006){Allard}, {Allard}, {Johnas}, \&
  {Kielkopf}}]{Cores06}
{Allard}, N.~F., {Allard}, F., {Johnas}, C., \& {Kielkopf}, J. 2006, 
  submitted to \aap 

\bibitem[{{Allard} {et~al.}(2005){Allard}, {Allard}, \& {Kielkopf}}]{alkalisLi}
{Allard}, N.~F., {Allard}, F., \& {Kielkopf}, J.~F. 2005, \aap, 440, 1195

\bibitem[{{Allard} {et~al.}(1999){Allard}, {Royer}, {Kielkopf}, \&
  {Feautrier}}]{NicolePhysRev99}
{Allard}, N.~F., {Royer}, A., {Kielkopf}, J., \& {Feautrier}, N. 1999, 
  \pra, 60, 1021

\bibitem[{{Burrows} {et~al.}(2000){Burrows}, {Marley}, \&
  {Sharp}}]{BurrowsMS00}
{Burrows}, A., {Marley}, M.~S., \& {Sharp}, C.~M. 2000, \apj, 531, 438

\bibitem[{{Dupuis} {et~al.}(1993){Dupuis}, {Fontaine}, \&
  {Wesemael}}]{dupuis93b}
{Dupuis}, J., {Fontaine}, G., \& {Wesemael}, F. 1993, \apjs, 87, 345

\bibitem[{{Harris} {et~al.}(2003){Harris}, {Liebert}, {Kleinman}, {Nitta},
  {Anderson}, {Knapp}, {Krzesi{\' n}ski}, {Schmidt}, {Strauss}, {Vanden Berk},
  {Eisenstein}, {Hawley}, {Margon}, {Munn}, {Silvestri}, {Smith}, {Szkody},
  {Collinge}, {Dahn}, {Fan}, {Hall}, {Schneider}, {Brinkmann}, {Burles},
  {Gunn}, {Hennessy}, {Hindsley}, {Ivezi{\' c}}, {Kent}, {Lamb}, {Lupton},
  {Nichol}, {Pier}, {Schlegel}, {SubbaRao}, {Uomoto}, {Yanny}, \&
  {York}}]{sdssWD03}
{Harris}, H.~C., {Liebert}, J., {Kleinman}, S.~J., {et~al.} 2003, \aj, 126,
  1023

\bibitem[{Hauschildt \& Baron(1999)}]{hbjcam99}
Hauschildt, P.~H. \& Baron, E. 1999, J. Comp. Applied Math., 109, 41

\bibitem[{{Homeier} {et~al.}(2005){Homeier}, {Allard}, {Allard}, {Hauschildt},
  {Schweitzer}, {Stancil}, \& {Weck}}]{alkaliWDkiel}
{Homeier}, D., {Allard}, N., {Allard}, F., {et~al.} 2005, in: 
  ASP Conf. Ser. 334: 14th European Workshop on White Dwarfs,
  eds. D.~Koester \& S.~Moehler (San Francisco), 209

\bibitem[{{Koester} {et~al.}(2005){Koester}, {Napiwotzki}, Voss, {Homeier}, \&
  {Reimers}}]{detlevDAZB05}
{Koester}, D., {Napiwotzki}, R., Voss, B., {Homeier}, D., \& {Reimers}, D.
  2005, \aap, 439, 317

\bibitem[Kowalski et al.(2005)]{2005ASPC..334..203K} Kowalski, P.~M.,
Saumon, D., \& Mazevet, S.\ 2005, ASP Conf.~Ser.~334, 203

\bibitem[{{Oppenheimer} {et~al.}(2001){Oppenheimer}, {Hambly}, {Digby},
  {Hodgkin}, \& {Saumon}}]{OHDHS01}
{Oppenheimer}, B.~R., {Hambly}, N.~C., {Digby}, A.~P., {Hodgkin}, S.~T., \&
  {Saumon}, D. 2001, Science, 292, 698

\bibitem[{{Salim} {et~al.}(2004){Salim}, {Rich}, {Hansen}, {Koopmans},
  {Oppenheimer}, \& {Blandford}}]{SalimHalo04}
{Salim}, S., {Rich}, R.~M., {Hansen}, B.~M., {et~al.} 2004, \apj, 601, 1075

\bibitem[{{Tsuji} {et~al.}(1999){Tsuji}, {Ohnaka}, \& {Aoki}}]{tsujiGl229}
{Tsuji}, T., {Ohnaka}, K., \& {Aoki}, W. 1999, \apjl, 520, L119

\bibitem[{{Wesemael} \& {Truran}(1982)}]{WT82}
{Wesemael}, F. \& {Truran}, J.~W. 1982, \apj, 260, 807

\end{thebibliography}

\end{document}